\documentclass[journal=jacsat,manuscript=article]{achemso}

\usepackage[version=3]{mhchem}

\usepackage[dvipsnames,svgnames,x11names]{xcolor}
\usepackage[markup=underlined]{changes}

\title{Example: changes}

\definechangesauthor[color=Blue]{JH}
\usepackage{todonotes}
\setcommentmarkup{\todo[color={authorcolor!20},size=\scriptsize]{#3: #1}}
\author{A.~O.~Mikhin}
\altaffiliation{Contributed equally to this work}
\author{A.~A.~Shubnic}
\altaffiliation{Contributed equally to this work}
\author{T.~V.~Ivanova}
\author{I.~A.~Shelykh}
\affiliation{School of Physics and Engineering, ITMO University, St. Petersburg, Russia}
\alsoaffiliation{Science Institute, University of Iceland, Dunhagi-3, IS-107, Reykjavik, Iceland}
\author{A.~K.~Samusev}
\author{I.~V.~Iorsh}
\affiliation{School of Physics and Engineering, ITMO University, St. Petersburg, Russia}
\email{i.iorsh@metalab.ifmo.ru}

\title[An \textsf{achemso} demo]
  {Bulk ReSe$_2$: record high refractive index and biaxially anisotropic material for all-dielectric nanophotonics}

\abbreviations{IR,NMR,UV}

\begin{tocentry}

\centering
\includegraphics[width=1\textwidth]{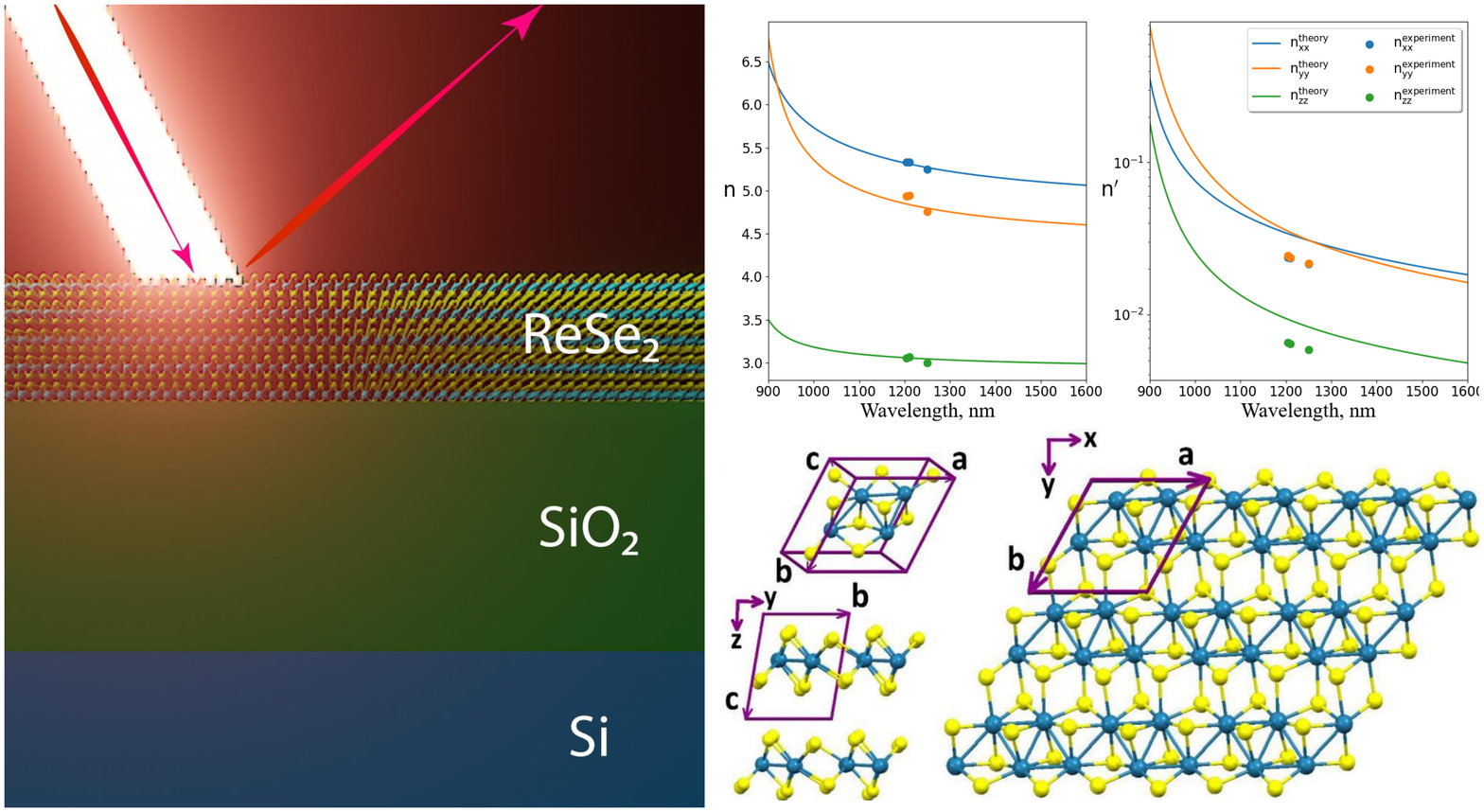}

\end{tocentry}

\begin{document}

\begin{abstract}
We show that bulk rhenium diselenide, ReSe$_2$ is characterized by record high value of the refractive index exceeding 5  in near-infrared frequency range. We use back focal plane reflection spectroscopy to extract the components of the  ReSe$_2$ permittivity tensor and reveal its extreme biaxial anisotropy. We also demonstrate the good agreement between the experimental data and the predictions of the density functional theory. The combination of the large refractive index and giant optical anisotropy makes ReSe$_2$ a perspective material for all-dielectric nanophotonics in the near-infrared frequency range.    
\end{abstract}

\section{Introduction}
The field of metamaterials and metasurfaces has recently witnessed the paradigm shift from plasmonic to all-dielectric nanophotonics~\cite{kuznetsov2016optically,kivshar2018all}. The main constitutive element of the latter is a subwavelength resonant nanoantenna which has two main advantages over its plasmonic counterpart: suppressed intrinsic Joule losses and the ability to support both electric and magnetic dipole Mie resonances of comparable strength~\cite{kuznetsov2012magnetic,evlyukhin2012}, which enables the achievement of high quality resonances in these subwavelength structures~\cite{koshelev2020}. All-dielectric metasurfaces allow to flexibly control both linear~\cite{decker2015high,arbabi2015dielectric,jahani2014all} and nonlinear~\cite{shcherbakov2015ultrafast, schlickriede2020nonlinear} optical properties, which makes possible their practical applications in the areas ranging from ultrathin abberation-free lenses~\cite{lin2014dielectric} to biosensing~\cite{jahani2021imaging}.   

The functionality of all-dielectric nanostructures is mainly limited by the value of the refractive index of a constituent material. In order to optimize the quality factor of the corresponding Mie resonances it is highly desirable to maximize the real part of the refractive index simultaneously minimize its imaginary part~\cite{Koshelev2018}. However, while there are plenty of materials characterized by extreme values of the refractive index ($n>10$) in the microwave, terahertz and even mid-infrared frequency range, in the near-infrared frequencies which are extremely important in terms of optical applications, such materials are lacking. As for now, the corresponding values of the refractive index are limited to $4.5$. This tendency is in line with the well known semi-empirical law~\cite{tripathy2015refractive}, which establishes the inverse scaling of a refractive index with a dielectric band gap, defining the transparency region of a material. At the same it is currently not clear whether for a given frequency there exists a fundamental upper limit for a value of the refractive index and the quest for finding high contrast optical compounds is on.

Recently it has been proposed theoretically~\cite{naccarato2019searching,rex2} that in search of high refractive index materials one could use a simple dimensionless parameter which can be straightforwardly extracted from the materials databases, namely the ratio of a band gap and a cumulative width of the conduction and valence bands. According to this criteria materials characterized by flat conduction and valence bands should have large refractive indices. In~\cite{rex2} we have found the specific material satisfying this criterion, namely rhenium diselenide (ReSe$_2$) and used the DFT-based first principles calculation to check this prediction and calculate the corresponding refractive index, which was shown to exceed 5 in a wide frequency range in near infrared. 

Rhenium diselenide corresponds to the family of bulk transition metal dichalcgonides (TMD). The optical properties of these materials have been extensively studied in the recent years. Specifically, it has been shown  that bulk MoS$_2$ is characterized by large value of refractive index and large anisotropy between in-plane and out-of plane components of dielectric permittivity tensors \cite{ermolaev2021giant}, which was followed by the measurements of optical constants for other TMD compounds~\cite{munkhbat2022optical}. Importantly, due to the uniaxial anisotropy of these Van der Waals materials, ellipsometric techniques is usually required for precise characterization of the dielectric permittivity tensor. In the case of the rhenium dichalcogenides~\cite{zhao2015interlayer,rahmanres2}, the situation is even more complicated since these compounds exhibit biaxial optical anisotropy. Recently, a method based on near-field probing of the waveguide dispersion has been used to the extract the dielectric permittivity tensor of ReS$_2$~\cite{res2anisotropy}.

In this Letter we realize an alternative method based on the angle-resolved reflection spectroscopy to measure all components of the dielectric permittivity tensor of an anisotropic crystal. We apply it for the case of ReSe$_2$ and show that the obtained results are in good agreement with the values obtained using DFT ab-initio calculation. It thus becomes confirmed that ReSe$_2$ is characterized by large biaxial anisotropy and extreme values of refractive index exceeding 5 in the wavelength range of 1200-1250 nm corresponding to near infrared .

\section{Results and discussion}

\begin{figure}[t!]
\centering
\includegraphics[width=1\textwidth]{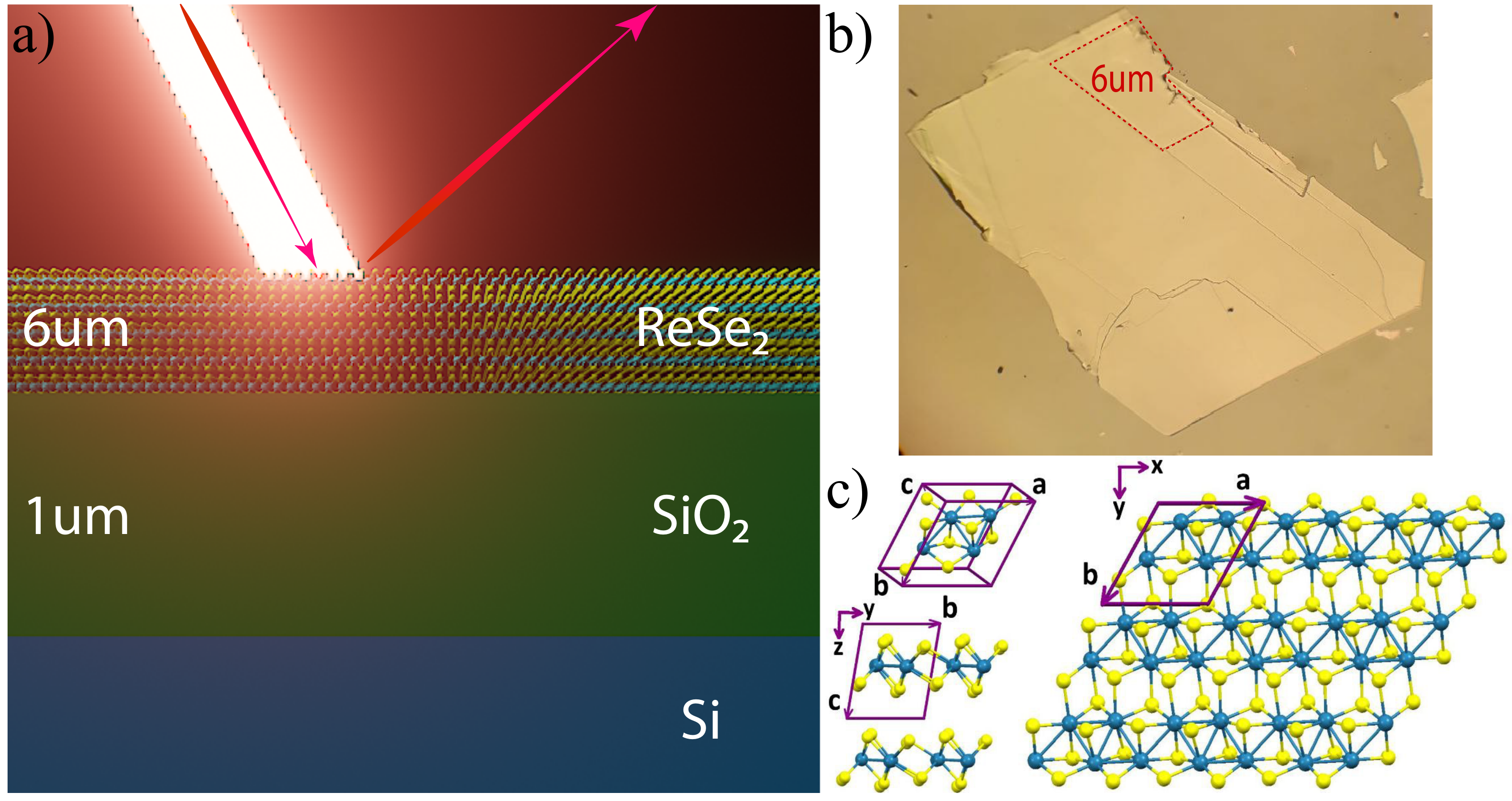}
\caption{\textbf{a)} Geometry of the experiment: we measured angle-resolved reflection spectra of a bulk ReSe$_2$ 6 um flake on 1 um SiO$_2$ layer on Si substrate. As a result, we obtained the dependence of the reflection coefficient on the projections of the wave vector \textbf{k} on the XY plane. The projections on the X and Y axes are designated as \textbf{k$_x$} and \textbf{k$_y$}, respectively. The XYZ coordinate system is shown in Fig. 1 c).\textbf{b)} Photograph of the sample. The dotted line indicates the area within which the thickness of the ReSe$_2$ is 6 um. \textbf{c)} Structure of ReSe$_2$ unit cell. It can be seen that ReSe$_2$ has a strongly anisotropic structure. Thus, it can be expected that the optical response of the layered structure considered in the experiment will also be strongly anisotropic.}
\label{fig:Sample_side_3D}
\end{figure}
The geometry of the experiment is shown in Fig.~\ref{fig:Sample_side_3D}. A thick flake of ReSe$_2$ was deposited on a thin film of SiO$_2$ placed on silicon substrate (See Methods). The lateral size of the flake was about 6~$\mu$m, which sufficiently exceeds the wavelength of the incident light. The flake thickness was determined by the atomic force microscopy measurements. It is well known that rhenium dichalcogenides are characterized by the skewed unit cell geometry as can be seen in the inset of Figure~\ref{fig:Sample_side_3D} which should lead to highly anisotropic optical response of these materials. 

In order to extract the anisotropic permittivity tensor we measure the angle-resolved reflectivity spectra. To take into account the in-plane anisotropy of the permittivity tensor, one needs to scan both polar and  azimuthal angles, and thus obtain the reflection coefficient profile in $(k_x,k_y)$ plane. We have used the back focal plane geometry of the experiment (see Methods) which allowed us to obtain the reflection spectra in almost whole $k_x,k_y$ plane at particular wavelength within a  single measurement. The measurement was performed in two orthogonal linear polarizations.

\begin{figure}[H]
\center{\includegraphics[width=0.9\linewidth]{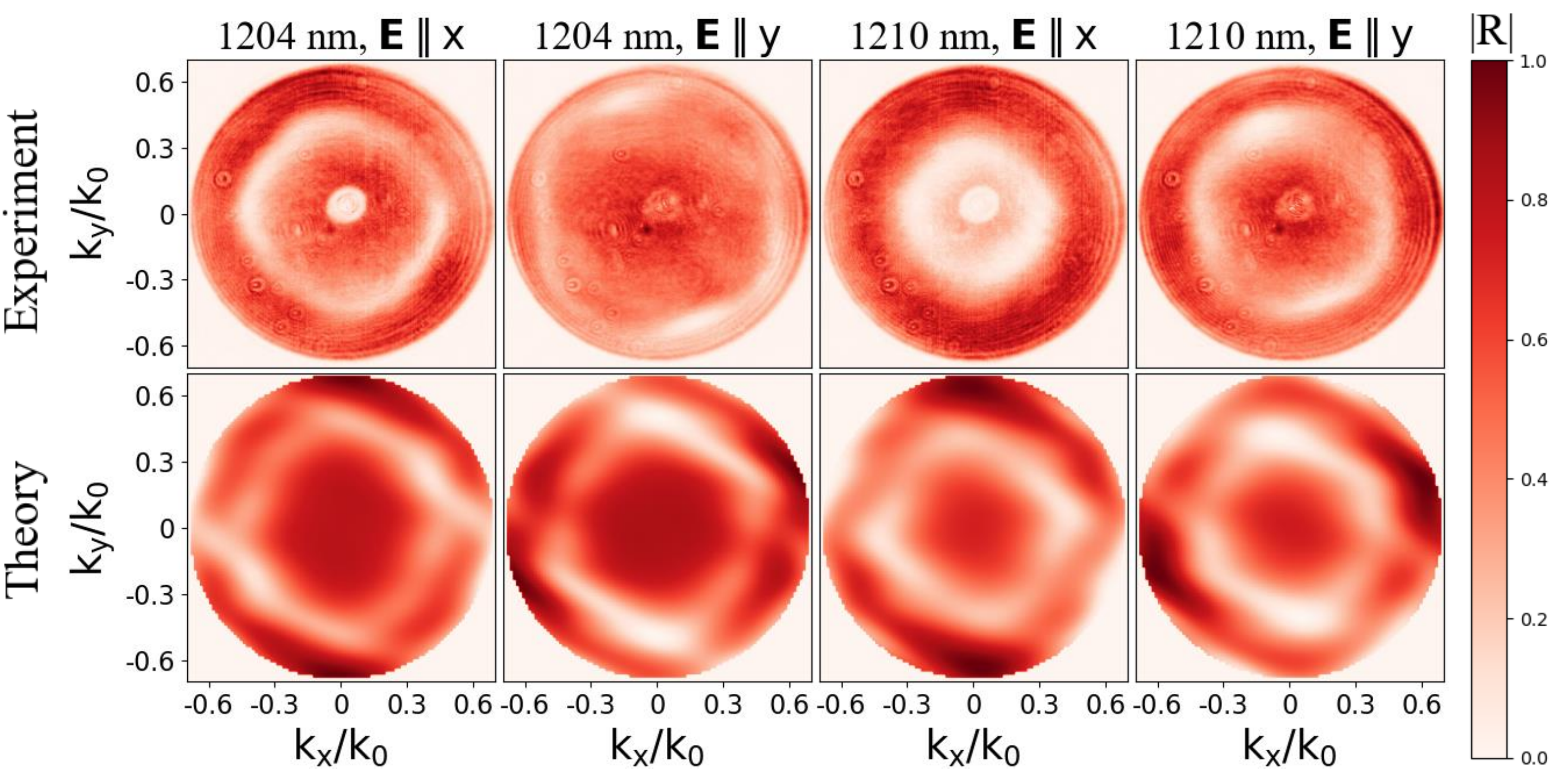}} 
\caption{Reflectivity profiles for wavelengths 1204 nm and 1210 nm and two perpendicular polarizations. The spectra are characterized by reflection minima forming a rhombus shape and arising from Fabry-Perot resonances in the layered structure. The top four dependencies are experimental, and the bottom four represent the result of modeling. In the presented simulation results, we used the dielectric permittivity tensor, which was obtained using machine learning-based optimization to best match the theoretical reflection spectra with experimental results. As an initial approximation of the optimization algorithm, the dielectric permittivity tensor obtained during the ab initio modeling was used. It can be seen that the theory reproduces both the shape and the average size of the experimental figures.} \label{Fig:2}
\end{figure}

The measured reflectivity profiles for the two wavelengths and two different polarizations are shown in the upper set of the images in Fig.~\ref{Fig:2}. One can clearly see the appearance of the non-circular contours of the reflectivity dips. They correspond to the formation of the Fabry-Perot resonances in the ReSe$_2$ slab. The average radius and the contrast of these contours depend crucially on the slab thickness: if the slab is too narrow, no Fabry-Perot resonances are formed, and if it is too thick, the losses suppress the contrast and smear out the reflectivity pattern. Therefore, in the course of the preliminary analysis, we have established the optimal thickness of the slab to be approximately equal to 6 $~\mu$m which we further used in our measurements. The observed complex, diamond-like  shape of the reflectivity dips contours distinct from both circular contours characteristic for isotropic materials and from elliptical ones appearing in the case of uniaxial crystals reflects the biaxial anisotropy of ReSe$_2$.

In order to extract the permittivity tensor we have obtained the analytical expression for the reflectivity in our experimental geometry using the generalized transfer matrix method (the details can be found in SI). We then performed numerical optimization procedure to fit the the dielectric permittivity tensor components in order to minimize the misfit of the theoretical and experimental reflectivity profiles for two orthogonal polarizations simultaneously. There was an additional global fitting parameter common for all the wavelengths: the angle between the principal component of the permittivity tensor and the plane of incidence. We have used the gradient optimization procedure with the dielectric permittivity tensor components obtained within the ab-initio calculations (see Methods) as the initial values. The theoretically calculated reflectivity profiles with the optimized parameters are shown in the lower set of images in Fig.~\ref{Fig:2}. One can see that the modelling reproduces the average size, shape, orientation and the linewidth of the experimental reflectivity dips. It should be noted that since we optimize not a single value of reflectivity at particular wavelength but the whole reflectivity map, this procedure results in accurate evaluation of the dielectric permittivity parameters.

\begin{figure}[H]
\center{\includegraphics[width=\linewidth]{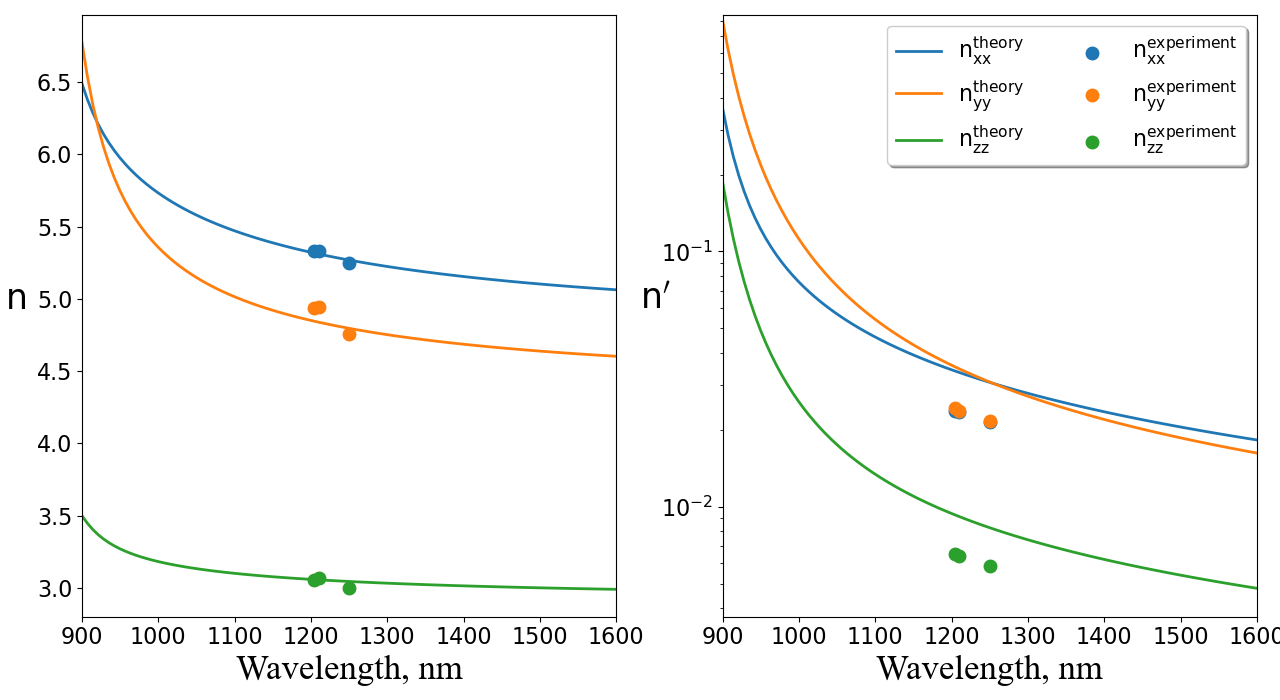}}
\caption{The values of the diagonal components of the refractive index tensor obtained in the first-principle modeling and extracted from the experiment. The experimental values were obtained as a result of machine learning-based optimization of the permittivity tensor for the best matching of the reflection spectra in Fig. 2. It can be seen that the theoretical and experimental results are in good agreement with each other.}\label{Fig:3}
\end{figure}

In Figure~\ref{Fig:3} we show the values of the three components of the real part of the refractive index obtained with DFT ab-initio methods (see Methods) and as optimal fit of the experimental data. We observe perfect correspondence between these values with the discrepancy not exceeding 5\%. 
The presented experimental results support our previous theoretical prediction that bulk ReSe$_2$ is characterized by  extreme biaxial anisotropy and record high value of the refractive index in the near infrared frequency range. In table I we present a comparison of the ReSe$_2$ refractive index with previously reported values for other compounds in the near-infrared frequency range. As to our knowledge, the refrective index of ReSe$_2$ in this diapason exceeds all previously reported values. 

\begin{table}[H]
  \label{tbl:example}
  \begin{tabular}{lllll}
    \hline
    Material  & $\lambda_{bound}$, nm  & n$_{max}$ & out-plane anisotropy & in-plane anisotropy\\
    \hline
    Si & 600 & 4.45 & - & -\\
    GaAs & 750 & 3.7 & - & -\\
    GaP & 450 & 3.9 & - & -\\
    Ge & 1100 & 4.3 & - & -\\
    MoS$_2$ & 750 & 4.8 & 1.5 & -\\
    h-BN & 450 & 2.2 & 0.8 & -\\
    MoTe$_2$ & 1270 & 5.2 & 2 & -\\
    ReS$_2$ & 900 & 4.3 & 1.7 & 0.35\\
    ReSe$_2$ & 900 & 5.5 & 2.4 & 0.5\\
    \hline
  \end{tabular}
  \caption{Comparison of refractive indices of ReSe$_2$ and previously reported high-index compounds. It can be seen that refractive index of ReSe$_2$ in the near-infrared frequency range exceeds refractive index of all previously reported materials. Also, ReSe$_2$ has excellent values of anisotropy both in and out of the plane.}
\end{table}

\subsection{Conclusions}

In conclusion, we applied the method of back focal plane reflection spectroscopy to obtain the spectrum of the dielectric permittivity tensor of the bulk biaxially anisotropic material ReSe$_2$. Our results demonstrate the record high value of the refractive index for this compound in the near-infrared spectral range. Combined with giant optical anisotropy, it makes ReSe$_2$ an ideal candidate for all-dielectric nanophotonics applications.

The work was supported by Ministry of Science and Higher Education of Russian Federation, goszadanie no. 2019-1246 and Priority 2030 Federal Academic Leadership Program. I.A.S. and I.V.I acknowledge the support from the joint RFBR-DFG project No. 21-52-12038.

\subsection{Theoretical methods}

Ab initio calculations were performed within the frameworks of DFT using the QUANTUM ESPRESSO package.~\cite{giannozzi09} We used PBE pseudopotentials~\cite{pbe96} from PSLibrary~\cite{PAW_base} in all calculations. 

In the first stage, the spatial structure of the unit cell was fully optimized using XRD experimental coordinates as a zero approximation.~\cite{XRD} The Broyden-Fletcher-Goldfarb-Shanno algorithm (BFGS) and Monkhorst-Pack method of special points with an unbiased 4x4x4 grid were used to determine the equilibrium configuration. The convergence threshold for self consistency was taken $10^{-10}$ Rydberg. Relativistic PAW pseudopotentials~\cite{PAW} were used.

In the second stage a self-consistent calculation of the electronic structure based on Broyden algorithm with non-relativistic USPP pseudopotentials~\cite{USPP} was carried out. All occupied orbitals were accounted for (188 electrons on 94 orbitals), as well as 36 unoccupied orbitals.

Finally, the optical response within the TDDFT~\cite{TDDFT} was determined based on a self-consistent calculation in $\Gamma$ point. The broadening/damping parameter was taken to be 25 meV.

Modeling of the reflection coefficient of a layered structure containing a slab of $ReSe_2$ was performed with use of the Berreman 4x4 matrix method~\cite{berreman1972} (the details can be found in SI).

\subsection{Experimental methods}

Sheets of rhenium diselenide was obtained from bulk crystal using mechanical exfoliation technique.
For the transferring of $\mathrm{Re Se_2}$ layers on the substrate we used commercial polydimethylsiloxane (PDMS) stamp. 
To avoid contamination, pristine exfoliated $\mathrm{Re Se_2}$ sheets was immediately transferred to the $\mathrm{Si O_2}$ substrate preliminarily cleaned with acetone and IPA. 
For better transfer of large layers to the substrate, we used a heavy press of 5 kg for 10 min. 
The next step was to characterize the bulk crystal thickness using atomic force microscopy (AFM) and find plane area of the $\mathrm{Re Se_2}$ sheet with thickness of 6 µm. 
At this thickness, the simulation showed the presence of well defined Fabry–Pérot resonances.
\begin{figure}[t!]
\centering
\includegraphics[width=1\textwidth]{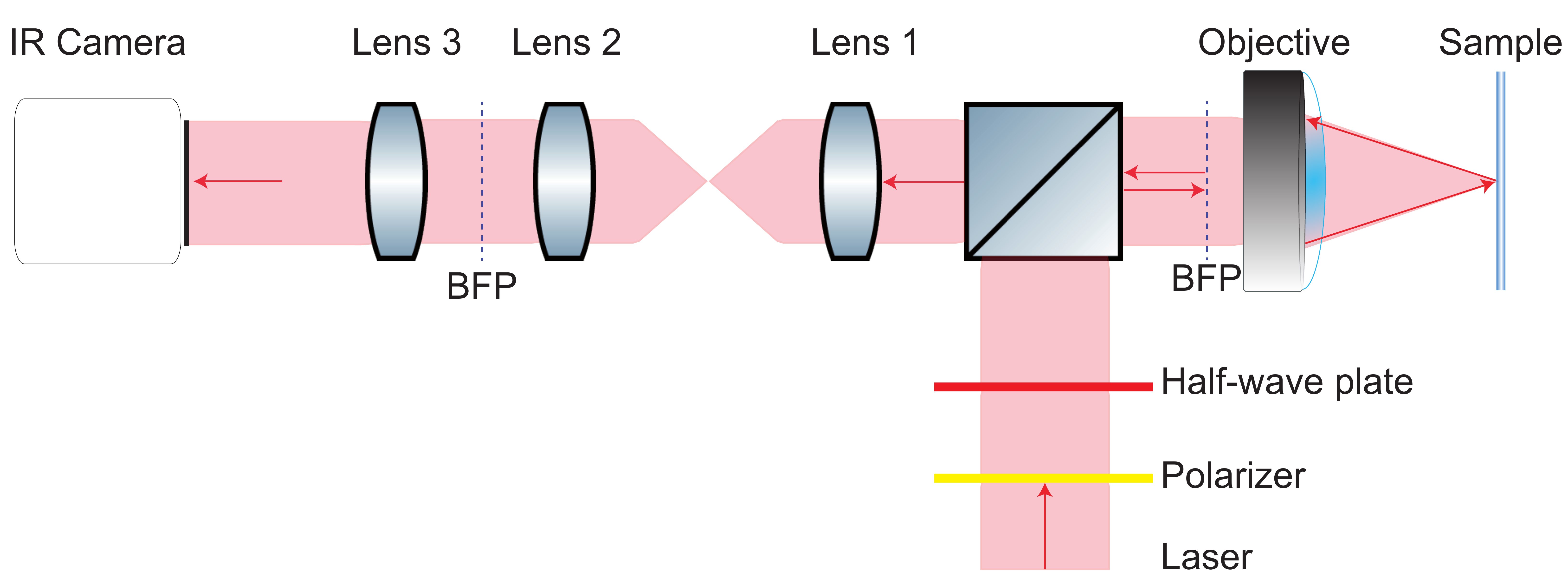}
\caption{Schematic view of the IR back focal plane setup. Red arrows shows the direction of the light propagation. To rotate the polarization of the laser injection used the Half-wave plate. The light passed through it focus on the sample by the objective. Reflectance signal collected through the same objective was recorded by the IR camera.}
\label{fig:IR_BFP_scheme}
\end{figure}
In order to determine a refractive index of the $\mathrm{Re Se_2}$ we carried out angle resolved spectroscopy measurements and obtained the isofrequency spectra.
We provided the experimental investigation in the reflection geometry, a scheme of our experimental setup is shown in Fig.\ref{fig:IR_BFP_scheme}. 
We used a femtosecond laser pulses provided by the optical parametric amplifier (Light Conversion Orpheus) pumped by 250 fs Yb laser (LightConversion Pharos) at the repetition rate of 1 MHz. 
In order to get angle resolved reflectance images of the sample at different wavelengths we tuned the wavelength using Light Conversion Orpheus in the range of the wavelength setting from 1017 to 3000 nm.

The spectral range of the laser pulse was about 20 nm, which is wider, than what is required for our experiment,  
Therefore, to narrow down the laser pulse spectral width to 1 nm we implemented the scheme for a spectral constriction.~\cite{miller2011probe} 

To control the polarization of the excitation laser we used a Glan-Taylor Polarizer that was placed before the setup.
To rotate the polarization plane we used a half-wave plate.

For focusing of the laser light we used Mitutoyo objective with magnification 100x and numerical aperture 0.7.
The reflectance signal was collected by the same objective and was visualized in the infrared CCD camera.

\subsection{Supporting Information}
Supporting Information Available: Experimental and simulation results for the reflection profile at a wavelength of 1250 nm and details of layered structure modeling.

\bibliography{bib-file}

\providecommand{\latin}[1]{#1}
\makeatletter
\providecommand{\doi}
  {\begingroup\let\do\@makeother\dospecials
  \catcode`\{=1 \catcode`\}=2 \doi@aux}
\providecommand{\doi@aux}[1]{\endgroup\texttt{#1}}
\makeatother
\providecommand*\mcitethebibliography{\thebibliography}
\csname @ifundefined\endcsname{endmcitethebibliography}
  {\let\endmcitethebibliography\endthebibliography}{}
\begin{mcitethebibliography}{31}
\providecommand*\natexlab[1]{#1}
\providecommand*\mciteSetBstSublistMode[1]{}
\providecommand*\mciteSetBstMaxWidthForm[2]{}
\providecommand*\mciteBstWouldAddEndPuncttrue
  {\def\EndOfBibitem{\unskip.}}
\providecommand*\mciteBstWouldAddEndPunctfalse
  {\let\EndOfBibitem\relax}
\providecommand*\mciteSetBstMidEndSepPunct[3]{}
\providecommand*\mciteSetBstSublistLabelBeginEnd[3]{}
\providecommand*\EndOfBibitem{}
\mciteSetBstSublistMode{f}
\mciteSetBstMaxWidthForm{subitem}{(\alph{mcitesubitemcount})}
\mciteSetBstSublistLabelBeginEnd
  {\mcitemaxwidthsubitemform\space}
  {\relax}
  {\relax}

\bibitem[Kuznetsov \latin{et~al.}(2016)Kuznetsov, Miroshnichenko, Brongersma,
  Kivshar, and Luk’yanchuk]{kuznetsov2016optically}
Kuznetsov,~A.~I.; Miroshnichenko,~A.~E.; Brongersma,~M.~L.; Kivshar,~Y.~S.;
  Luk’yanchuk,~B. Optically resonant dielectric nanostructures.
  \emph{Science} \textbf{2016}, \emph{354}\relax
\mciteBstWouldAddEndPuncttrue
\mciteSetBstMidEndSepPunct{\mcitedefaultmidpunct}
{\mcitedefaultendpunct}{\mcitedefaultseppunct}\relax
\EndOfBibitem
\bibitem[Kivshar(2018)]{kivshar2018all}
Kivshar,~Y. All-dielectric meta-optics and non-linear nanophotonics.
  \emph{National Science Review} \textbf{2018}, \emph{5}, 144--158\relax
\mciteBstWouldAddEndPuncttrue
\mciteSetBstMidEndSepPunct{\mcitedefaultmidpunct}
{\mcitedefaultendpunct}{\mcitedefaultseppunct}\relax
\EndOfBibitem
\bibitem[Kuznetsov \latin{et~al.}(2012)Kuznetsov, Miroshnichenko, Fu, Zhang,
  and Luk’Yanchuk]{kuznetsov2012magnetic}
Kuznetsov,~A.~I.; Miroshnichenko,~A.~E.; Fu,~Y.~H.; Zhang,~J.;
  Luk’Yanchuk,~B. Magnetic light. \emph{Scientific reports} \textbf{2012},
  \emph{2}, 492\relax
\mciteBstWouldAddEndPuncttrue
\mciteSetBstMidEndSepPunct{\mcitedefaultmidpunct}
{\mcitedefaultendpunct}{\mcitedefaultseppunct}\relax
\EndOfBibitem
\bibitem[Evlyukhin \latin{et~al.}(2012)Evlyukhin, Novikov, Zywietz, Eriksen,
  Reinhardt, Bozhevolnyi, and Chichkov]{evlyukhin2012}
Evlyukhin,~A.~B.; Novikov,~S.~M.; Zywietz,~U.; Eriksen,~R.~L.; Reinhardt,~C.;
  Bozhevolnyi,~S.~I.; Chichkov,~B.~N. Demonstration of magnetic dipole
  resonances of dielectric nanospheres in the visible region. \emph{Nano
  Letters} \textbf{2012}, \emph{12}, 3749–3755\relax
\mciteBstWouldAddEndPuncttrue
\mciteSetBstMidEndSepPunct{\mcitedefaultmidpunct}
{\mcitedefaultendpunct}{\mcitedefaultseppunct}\relax
\EndOfBibitem
\bibitem[Koshelev \latin{et~al.}(2020)Koshelev, Kruk, Melik-Gaykazyan, Choi,
  Bogdanov, Park, and Kivshar]{koshelev2020}
Koshelev,~K.; Kruk,~S.; Melik-Gaykazyan,~E.; Choi,~J.-H.; Bogdanov,~A.;
  Park,~H.-G.; Kivshar,~Y. Subwavelength dielectric resonators for nonlinear
  nanophotonics. \emph{Science} \textbf{2020}, \emph{367}, 288--292\relax
\mciteBstWouldAddEndPuncttrue
\mciteSetBstMidEndSepPunct{\mcitedefaultmidpunct}
{\mcitedefaultendpunct}{\mcitedefaultseppunct}\relax
\EndOfBibitem
\bibitem[Decker \latin{et~al.}(2015)Decker, Staude, Falkner, Dominguez, Neshev,
  Brener, Pertsch, and Kivshar]{decker2015high}
Decker,~M.; Staude,~I.; Falkner,~M.; Dominguez,~J.; Neshev,~D.~N.; Brener,~I.;
  Pertsch,~T.; Kivshar,~Y.~S. High-efficiency dielectric Huygens’ surfaces.
  \emph{Advanced Optical Materials} \textbf{2015}, \emph{3}, 813--820\relax
\mciteBstWouldAddEndPuncttrue
\mciteSetBstMidEndSepPunct{\mcitedefaultmidpunct}
{\mcitedefaultendpunct}{\mcitedefaultseppunct}\relax
\EndOfBibitem
\bibitem[Arbabi \latin{et~al.}(2015)Arbabi, Horie, Bagheri, and
  Faraon]{arbabi2015dielectric}
Arbabi,~A.; Horie,~Y.; Bagheri,~M.; Faraon,~A. Dielectric metasurfaces for
  complete control of phase and polarization with subwavelength spatial
  resolution and high transmission. \emph{Nature nanotechnology} \textbf{2015},
  \emph{10}, 937--943\relax
\mciteBstWouldAddEndPuncttrue
\mciteSetBstMidEndSepPunct{\mcitedefaultmidpunct}
{\mcitedefaultendpunct}{\mcitedefaultseppunct}\relax
\EndOfBibitem
\bibitem[Jahani and Jacob(2014)Jahani, and Jacob]{jahani2014all}
Jahani,~S.; Jacob,~Z. All-dielectric metamaterials. \emph{Nature
  Nanotechnology} \textbf{2014}, \emph{11}, 23--36\relax
\mciteBstWouldAddEndPuncttrue
\mciteSetBstMidEndSepPunct{\mcitedefaultmidpunct}
{\mcitedefaultendpunct}{\mcitedefaultseppunct}\relax
\EndOfBibitem
\bibitem[Shcherbakov \latin{et~al.}(2015)Shcherbakov, Vabishchevich, Shorokhov,
  Chong, Choi, Staude, Miroshnichenko, Neshev, Fedyanin, and
  Kivshar]{shcherbakov2015ultrafast}
Shcherbakov,~M.~R.; Vabishchevich,~P.~P.; Shorokhov,~A.~S.; Chong,~K.~E.;
  Choi,~D.-Y.; Staude,~I.; Miroshnichenko,~A.~E.; Neshev,~D.~N.;
  Fedyanin,~A.~A.; Kivshar,~Y.~S. Ultrafast all-optical switching with magnetic
  resonances in nonlinear dielectric nanostructures. \emph{Nano letters}
  \textbf{2015}, \emph{15}, 6985--6990\relax
\mciteBstWouldAddEndPuncttrue
\mciteSetBstMidEndSepPunct{\mcitedefaultmidpunct}
{\mcitedefaultendpunct}{\mcitedefaultseppunct}\relax
\EndOfBibitem
\bibitem[Schlickriede \latin{et~al.}(2020)Schlickriede, Kruk, Wang, Sain,
  Kivshar, and Zentgraf]{schlickriede2020nonlinear}
Schlickriede,~C.; Kruk,~S.~S.; Wang,~L.; Sain,~B.; Kivshar,~Y.; Zentgraf,~T.
  Nonlinear imaging with all-dielectric metasurfaces. \emph{Nano Letters}
  \textbf{2020}, \relax
\mciteBstWouldAddEndPunctfalse
\mciteSetBstMidEndSepPunct{\mcitedefaultmidpunct}
{}{\mcitedefaultseppunct}\relax
\EndOfBibitem
\bibitem[Lin \latin{et~al.}(2014)Lin, Fan, Hasman, and
  Brongersma]{lin2014dielectric}
Lin,~D.; Fan,~P.; Hasman,~E.; Brongersma,~M.~L. Dielectric gradient metasurface
  optical elements. \emph{science} \textbf{2014}, \emph{345}, 298--302\relax
\mciteBstWouldAddEndPuncttrue
\mciteSetBstMidEndSepPunct{\mcitedefaultmidpunct}
{\mcitedefaultendpunct}{\mcitedefaultseppunct}\relax
\EndOfBibitem
\bibitem[Jahani \latin{et~al.}(2021)Jahani, Arvelo, Yesilkoy, Koshelev,
  Cianciaruso, De~Palma, Kivshar, and Altug]{jahani2021imaging}
Jahani,~Y.; Arvelo,~E.~R.; Yesilkoy,~F.; Koshelev,~K.; Cianciaruso,~C.;
  De~Palma,~M.; Kivshar,~Y.; Altug,~H. Imaging-based spectrometer-less
  optofluidic biosensors based on dielectric metasurfaces for detecting
  extracellular vesicles. \emph{Nature Communications} \textbf{2021},
  \emph{12}, 1--10\relax
\mciteBstWouldAddEndPuncttrue
\mciteSetBstMidEndSepPunct{\mcitedefaultmidpunct}
{\mcitedefaultendpunct}{\mcitedefaultseppunct}\relax
\EndOfBibitem
\bibitem[Koshelev \latin{et~al.}(2018)Koshelev, Lepeshov, Liu, Bogdanov, and
  Kivshar]{Koshelev2018}
Koshelev,~K.; Lepeshov,~S.; Liu,~M.; Bogdanov,~A.; Kivshar,~Y. Asymmetric
  metasurfaces with high-Q resonances governed by bound states in the
  continuum. \emph{Physical review letters} \textbf{2018}, \emph{121},
  193903\relax
\mciteBstWouldAddEndPuncttrue
\mciteSetBstMidEndSepPunct{\mcitedefaultmidpunct}
{\mcitedefaultendpunct}{\mcitedefaultseppunct}\relax
\EndOfBibitem
\bibitem[Tripathy(2015)]{tripathy2015refractive}
Tripathy,~S. Refractive indices of semiconductors from energy gaps.
  \emph{Optical materials} \textbf{2015}, \emph{46}, 240--246\relax
\mciteBstWouldAddEndPuncttrue
\mciteSetBstMidEndSepPunct{\mcitedefaultmidpunct}
{\mcitedefaultendpunct}{\mcitedefaultseppunct}\relax
\EndOfBibitem
\bibitem[Naccarato \latin{et~al.}(2019)Naccarato, Ricci, Suntivich, Hautier,
  Wirtz, and Rignanese]{naccarato2019searching}
Naccarato,~F.; Ricci,~F.; Suntivich,~J.; Hautier,~G.; Wirtz,~L.;
  Rignanese,~G.-M. Searching for materials with high refractive index and wide
  band gap: A first-principles high-throughput study. \emph{Physical Review
  Materials} \textbf{2019}, \emph{3}, 044602\relax
\mciteBstWouldAddEndPuncttrue
\mciteSetBstMidEndSepPunct{\mcitedefaultmidpunct}
{\mcitedefaultendpunct}{\mcitedefaultseppunct}\relax
\EndOfBibitem
\bibitem[Shubnic \latin{et~al.}(2020)Shubnic, Polozkov, Shelykh, and
  Iorsh]{rex2}
Shubnic,~A.~A.; Polozkov,~R.~G.; Shelykh,~I.~A.; Iorsh,~I.~V. High refractive
  index and extreme biaxial optical anisotropy of rhenium diselenide for
  applications in all-dielectric nanophotonics. \emph{Nanophotonics}
  \textbf{2020}, \emph{9}, 4737--4742\relax
\mciteBstWouldAddEndPuncttrue
\mciteSetBstMidEndSepPunct{\mcitedefaultmidpunct}
{\mcitedefaultendpunct}{\mcitedefaultseppunct}\relax
\EndOfBibitem
\bibitem[Ermolaev \latin{et~al.}(2021)Ermolaev, Grudinin, Stebunov, Voronin,
  Kravets, Duan, Mazitov, Tselikov, Bylinkin, Yakubovsky, \latin{et~al.}
  others]{ermolaev2021giant}
Ermolaev,~G.; Grudinin,~D.; Stebunov,~Y.; Voronin,~K.~V.; Kravets,~V.;
  Duan,~J.; Mazitov,~A.; Tselikov,~G.; Bylinkin,~A.; Yakubovsky,~D.,
  \latin{et~al.}  Giant optical anisotropy in transition metal dichalcogenides
  for next-generation photonics. \emph{Nature communications} \textbf{2021},
  \emph{12}, 1--8\relax
\mciteBstWouldAddEndPuncttrue
\mciteSetBstMidEndSepPunct{\mcitedefaultmidpunct}
{\mcitedefaultendpunct}{\mcitedefaultseppunct}\relax
\EndOfBibitem
\bibitem[Munkhbat \latin{et~al.}(2022)Munkhbat, Wr{\'o}bel, Antosiewicz, and
  Shegai]{munkhbat2022optical}
Munkhbat,~B.; Wr{\'o}bel,~P.; Antosiewicz,~T.~J.; Shegai,~T.~O. Optical
  Constants of Several Multilayer Transition Metal Dichalcogenides Measured by
  Spectroscopic Ellipsometry in the 300--1700 nm Range: High Index, Anisotropy,
  and Hyperbolicity. \emph{ACS Photonics} \textbf{2022}, \relax
\mciteBstWouldAddEndPunctfalse
\mciteSetBstMidEndSepPunct{\mcitedefaultmidpunct}
{}{\mcitedefaultseppunct}\relax
\EndOfBibitem
\bibitem[Zhao \latin{et~al.}(2015)Zhao, Wu, Zhong, Guo, Wang, Xia, Yang, Tan,
  and Wang]{zhao2015interlayer}
Zhao,~H.; Wu,~J.; Zhong,~H.; Guo,~Q.; Wang,~X.; Xia,~F.; Yang,~L.; Tan,~P.;
  Wang,~H. Interlayer interactions in anisotropic atomically thin rhenium
  diselenide. \emph{Nano Research} \textbf{2015}, \emph{8}, 3651--3661\relax
\mciteBstWouldAddEndPuncttrue
\mciteSetBstMidEndSepPunct{\mcitedefaultmidpunct}
{\mcitedefaultendpunct}{\mcitedefaultseppunct}\relax
\EndOfBibitem
\bibitem[Rahman \latin{et~al.}(2017)Rahman, Davey, and Qiao]{rahmanres2}
Rahman,~M.; Davey,~K.; Qiao,~S.-Z. Advent of 2D rhenium disulfide (ReS2):
  fundamentals to applications. \emph{Advanced Functional Materials}
  \textbf{2017}, \emph{27}, 1606129\relax
\mciteBstWouldAddEndPuncttrue
\mciteSetBstMidEndSepPunct{\mcitedefaultmidpunct}
{\mcitedefaultendpunct}{\mcitedefaultseppunct}\relax
\EndOfBibitem
\bibitem[Mooshammer \latin{et~al.}(2022)Mooshammer, Chae, Zhang, Shao, Qiu,
  Rajendran, Sternbach, Rizzo, Zhu, Schuck, Hone, and Basov]{res2anisotropy}
Mooshammer,~F.; Chae,~S.; Zhang,~S.; Shao,~Y.; Qiu,~S.; Rajendran,~A.;
  Sternbach,~A.~J.; Rizzo,~D.~J.; Zhu,~X.; Schuck,~P.~J.; Hone,~J.~C.;
  Basov,~D.~N. In-plane anisotropy in biaxial ReS2 crystals probed by
  nano-optical imaging of waveguide modes. \emph{ACS Photonics} \textbf{2022},
  \emph{9}, 443--451\relax
\mciteBstWouldAddEndPuncttrue
\mciteSetBstMidEndSepPunct{\mcitedefaultmidpunct}
{\mcitedefaultendpunct}{\mcitedefaultseppunct}\relax
\EndOfBibitem
\bibitem[Giannozzi \latin{et~al.}(2009)Giannozzi, Baroni, Bonini, Calandra,
  Car, Cavazzoni, Ceresoli, Chiarotti, Cococcioni, Dabo, Corso, de~Gironcoli,
  Fabris, Fratesi, Gebauer, Gerstmann, Gougoussis, Kokalj, Lazzeri,
  Martin-Samos, Marzari, Mauri, Mazzarello, Paolini, Pasquarello, Paulatto,
  Sbraccia, Scandolo, Sclauzero, Seitsonen, Smogunov, Umari, and
  Wentzcovitch]{giannozzi09}
Giannozzi,~P. \latin{et~al.}  QUANTUM ESPRESSO: a modular and open-source
  software project for quantum simulations of materials. \emph{Journal of
  Physics: Condensed Matter} \textbf{2009}, \emph{21}, 395502\relax
\mciteBstWouldAddEndPuncttrue
\mciteSetBstMidEndSepPunct{\mcitedefaultmidpunct}
{\mcitedefaultendpunct}{\mcitedefaultseppunct}\relax
\EndOfBibitem
\bibitem[Perdew \latin{et~al.}(1996)Perdew, Burke, and Ernzerhof]{pbe96}
Perdew,~J.~P.; Burke,~K.; Ernzerhof,~M. Generalized Gradient Approximation Made
  Simple. \emph{Phys. Rev. Lett.} \textbf{1996}, \emph{77}, 3865--3868\relax
\mciteBstWouldAddEndPuncttrue
\mciteSetBstMidEndSepPunct{\mcitedefaultmidpunct}
{\mcitedefaultendpunct}{\mcitedefaultseppunct}\relax
\EndOfBibitem
\bibitem[Corso(2014)]{PAW_base}
Corso,~A.~D. Pseudopotentials periodic table: From H to Pu. \emph{Comput.
  Mater. Sci.} \textbf{2014}, \emph{95}, 337--350\relax
\mciteBstWouldAddEndPuncttrue
\mciteSetBstMidEndSepPunct{\mcitedefaultmidpunct}
{\mcitedefaultendpunct}{\mcitedefaultseppunct}\relax
\EndOfBibitem
\bibitem[Lamfers \latin{et~al.}(1996)Lamfers, Meetsma, Wiegers, and
  deBoer]{XRD}
Lamfers,~H.; Meetsma,~A.; Wiegers,~G.; deBoer,~J. The crystal structure of some
  rhenium and technetium dichalcogenides. \emph{Journal of Alloys and
  Compounds} \textbf{1996}, \emph{241}, 34 -- 39\relax
\mciteBstWouldAddEndPuncttrue
\mciteSetBstMidEndSepPunct{\mcitedefaultmidpunct}
{\mcitedefaultendpunct}{\mcitedefaultseppunct}\relax
\EndOfBibitem
\bibitem[Kresse and Joubert(1999)Kresse, and Joubert]{PAW}
Kresse,~G.; Joubert,~D. From ultrasoft pseudopotentials to the projector
  augmented-wave method. \emph{Phys. Rev. B} \textbf{1999}, \emph{59},
  1758--1775\relax
\mciteBstWouldAddEndPuncttrue
\mciteSetBstMidEndSepPunct{\mcitedefaultmidpunct}
{\mcitedefaultendpunct}{\mcitedefaultseppunct}\relax
\EndOfBibitem
\bibitem[Vanderbilt(1990)]{USPP}
Vanderbilt,~D. Soft self-consistent pseudopotentials in a generalized
  eigenvalue formalism. \emph{Phys. Rev. B} \textbf{1990}, \emph{41},
  7892\relax
\mciteBstWouldAddEndPuncttrue
\mciteSetBstMidEndSepPunct{\mcitedefaultmidpunct}
{\mcitedefaultendpunct}{\mcitedefaultseppunct}\relax
\EndOfBibitem
\bibitem[Rocca \latin{et~al.}(2008)Rocca, Gebauer, Saad, and Baroni]{TDDFT}
Rocca,~D.; Gebauer,~R.; Saad,~Y.; Baroni,~S. Turbo charging time-dependent
  density-functional theory with Lanczos chains. \emph{The Journal of Chemical
  Physics} \textbf{2008}, \emph{128}, 154105\relax
\mciteBstWouldAddEndPuncttrue
\mciteSetBstMidEndSepPunct{\mcitedefaultmidpunct}
{\mcitedefaultendpunct}{\mcitedefaultseppunct}\relax
\EndOfBibitem
\bibitem[Berreman(1972)]{berreman1972}
Berreman,~D.~W. Optics in stratified and anisotropic media: 4× 4-matrix
  formulation. \emph{Journal of the Optical Society of America} \textbf{1972},
  \emph{62}, 502--510\relax
\mciteBstWouldAddEndPuncttrue
\mciteSetBstMidEndSepPunct{\mcitedefaultmidpunct}
{\mcitedefaultendpunct}{\mcitedefaultseppunct}\relax
\EndOfBibitem
\bibitem[Miller \latin{et~al.}(2011)Miller, Slipchenko, and
  Meyer]{miller2011probe}
Miller,~J.~D.; Slipchenko,~M.~N.; Meyer,~T.~R. Probe-pulse optimization for
  nonresonant suppression in hybrid fs/ps coherent anti-Stokes Raman scattering
  at high temperature. \emph{Optics Express} \textbf{2011}, \emph{19},
  13326--13333\relax
\mciteBstWouldAddEndPuncttrue
\mciteSetBstMidEndSepPunct{\mcitedefaultmidpunct}
{\mcitedefaultendpunct}{\mcitedefaultseppunct}\relax
\EndOfBibitem
\end{mcitethebibliography}

\end{document}


\section{Results for a wavelength of 1250 nm}

\begin{figure}[H]
\center{\includegraphics[width=0.61\linewidth]{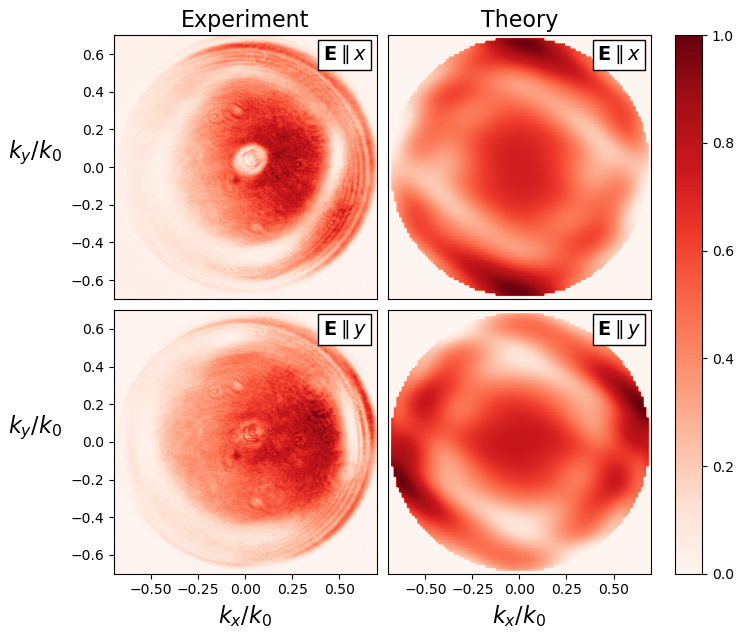}}
\caption{Comparison of reflectivity profiles in theory and experiment for both polarizations at the wavelength of 1250 nm. Optimized values were used as the real part of the permittivity tensor, while the imaginary part of the permittivity tensor was taken directly from the dft calculation.}
\end{figure}

\section{Details of layered structure modeling}

We used the 4x4 Berreman transfer matrix method. Maxwell's equations were written in matrix form in order to relate the longitudinal and tangential components of the electric and magnetic fields at the boundaries of an infinitesimal layer. This approach makes it possible to describe anisotropic media. Next, the exponent of the resulting matrix multiplied by $ik_0d$ was calculated, where $k_0$ is the wave vector of the incident wave in vacuum and $d$ is the thickness of the layer. Repeating this procedure for $ReSe_2$ and $SiO_2$ layers, matrices characterizing these layers were obtained. As a result of multiplying these matrices, a transfer matrix was obtained. The transfer matrix allows us to link incident, reflected and refracted light waves. Solving the resulting system of linear equations, we found the relations between the components of the electric and magnetic fields in the incident and refracted light waves and, using these relations, calculated the reflection coefficient.
